\documentclass[conference,a4paper]{IEEEtran}
\IEEEoverridecommandlockouts
\usepackage{cite}
\usepackage{amsmath,amssymb,amsfonts}
\usepackage{algorithmic}
\usepackage{graphicx}
\usepackage{textcomp}
\usepackage{xcolor}
\usepackage{stfloats}
\usepackage{subcaption}
\usepackage{pbalance}
\usepackage[font=footnotesize]{caption}
\usepackage{hyperref}
\hypersetup{
	colorlinks=true,
	linkcolor=black,
	citecolor=black,
	urlcolor=black
}

\def\BibTeX{{\rm B\kern-.05em{\sc i\kern-.025em b}\kern-.08em
    T\kern-.1667em\lower.7ex\hbox{E}\kern-.125emX}}
\begin{document}

\IEEEpubid{979-8-3315-2503-3/25/\$31.00~\copyright 2025 IEEE}

\title{Impact Assessment of Heterogeneous Grid Support Functions in Smart Inverter Deployments}

\author{
\IEEEauthorblockN{S. Gokul Krishnan, Mohd. Asim Aftab, Nabil Mohammed, Shehab Ahmed, Charalambos Konstantinou}

\IEEEauthorblockA{
CEMSE Division, King Abdullah University of Science and Technology (KAUST)}
}
\IEEEaftertitletext{\vspace{-0.4\baselineskip}}

\maketitle
\begin{abstract}
The decarbonization of the energy sector has led to a significant  high penetration of distributed energy resources (DERs), particularly photovoltaic (PV) systems, in low-voltage (LV) distribution networks.
To maintain grid stability, recent standards (e.g., IEEE 1547-2018)  mandate DERs to provide grid-support functionalities through smart inverters (SIs), which typically operate autonomously based on local measurements. 
However, as DER penetration increases, uncoordinated control modes of SIs can lead to adverse interactions, compromising system efficiency, voltage regulation, and overall stability. 
While previous studies have demonstrated the benefits of coordinated inverter control and optimal dispatch strategies, the system-wide impacts of heterogeneous SI groups operating under different control modes remain largely unexamined.  This paper addresses this gap by assessing the dynamic interactions among multiple SI groups with varying control strategies, namely: Constant Power Factor (CPF), Volt-VAR, and Volt-Watt modes. Furthermore, the analysis covers both resistive and inductive feeder types. 
The validation is performed using a real-time setup. The CIRGE low-voltage (LV) distribution network is simulated in the Opal-RT platform as the test network, enabling realistic and high-fidelity evaluation of SI control interactions under practical grid conditions.
\end{abstract}


\section{Introduction} \label{sec: intro}
The global shift toward decarbonization has accelerated the deployment of renewable energy resources, particularly photovoltaic (PV) systems \cite{renewable}. Initially, these distributed energy resources (DERs) were deployed with minimal coordination, as their low penetration posed negligible impact on grid stability. However, with the increasing proliferation of DERs, traditional grid operation and control paradigms have become insufficient. This necessitated the revision of IEEE 1547-2003 and the release of IEEE 1547-2018, which mandates essential grid support functionalities such as voltage, frequency ridethrough, power regulation, and ramp rate control of DERs \cite{IEEE1547}. These capabilities are facilitated by smart inverters (SIs), which are capable of dynamically adjusting their power output and can also communicate with the DER aggregators to enhance reliability and resilience.

SIs provide advanced ancillary services, including reactive power control (Volt-VAR), active power curtailment (Volt-Watt), frequency stabilization (Frequency-Watt), and ride-through capabilities under abnormal grid conditions \cite{EPRI}. The majority of SIs utilize preprogrammed control curves that define their active and reactive power responses based solely on local measurements obtained via phase-locked loops (PLLs).  This decentralized control architecture enables autonomous inverter operation without requiring a centralized communication framework. However, the control behavior of SIs is highly dependent on their electrical and geographical deployment context \cite{Coordination}. The selection of SIs operating mode depends on factors such as feeder location/topology, local voltage level, power quality metrics, and utility interconnection standards. For instance, an inverter located at the end of a radial distribution feeder may operate in Volt-Var mode for local voltage support, whereas one near a substation may be configured to maintain a constant power  (CPF). Distribution system operators (DSOs) may mandate specific control modes to address feeder-specific constraints or mitigate grid disturbances.

As DER penetration continues to increase, the uncoordinated, locally governed operation of SIs can lead to adverse interactions. Since each SI operates independently based on its own local measurements, their collective response may result in control conflicts, reduced voltage regulation effectiveness, or decreased overall system efficiency and stability. Although coordinated PV inverter control strategies have demonstrated benefits such as reduced power curtailment, lower line losses, improved PV hosting capacity, and enhanced system stability \cite{AddedValueofCoordination}, most existing studies assume homogeneous inverter operation or rely on centralized optimization frameworks \cite{OID1,OID2}. The system-level implications of heterogeneous control mode deployment—where different groups of SIs operate under distinct control strategies—remain largely unexplored. In particular, the cross-impact of control mode changes in one SI group on the dynamic performance of other groups within the network has not yet been systematically investigated.

\IEEEpubidadjcol

To address this research gap, this paper presents a comprehensive impact assessment of heterogeneous grid support functions deployed via SIs in a LV distribution network.
A real-time simulation framework is employed using the Opal-RT simulator to model the CIGRE LV distribution network under both resistive and inductive line configurations. 
This high-fidelity platform enables realistic evaluation and validation of the dynamic interactions among SI groups operating under heterogeneous control modes. This study contributes novel insights into the collective dynamics of SIs with heterogeneous control modes and demonstrates their influence on voltage regulation in distribution networks.

The rest of the paper is organized as follows. Section~\ref{sec:system_desc} describes the electric grid architecture and functions of SI. Section~\ref{sec:system_modeling} details the system modeling used for the analysis. Section~\ref{sec:results} presents the simulation results. Finally, Section~\ref{sec:conclusion} concludes the paper.


\section{System Description and Control of SIs}\label{sec:system_desc}
This section outlines the fundamental principles governing the relationship between power injections and voltage behavior in electric power systems, with emphasis on the influence of line impedance characteristics. It further introduces widely adopted control strategies of SIs for voltage regulation in low- and medium-voltage networks.

\subsection{Fundamentals of Electric Grid Modeling}
The coupling between power flows and voltage profiles in electric networks is strongly influenced by the impedance characteristics of the lines. In high-voltage (HV) transmission networks, where line reactance dominates resistance $(\frac{X}{R}>>1)$, voltage variations are primarily influenced by reactive power flows. In contrast, LV distribution networks exhibit a  higher resistance-to-reactance ratios $(\frac{R}{X}>>1)$, where voltage variations are more sensitive to active power injections \cite{twobus}. 
\begin{figure}[t]
    \centering
    \includegraphics[scale=0.3]{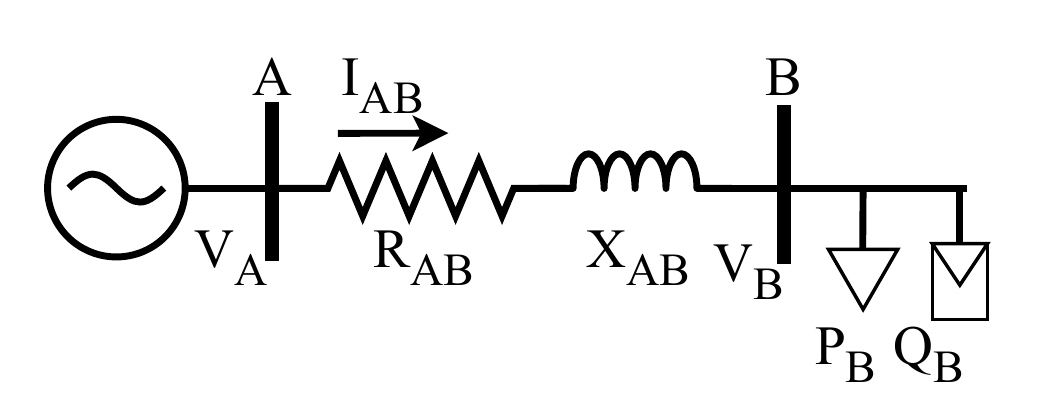}
    \vspace{-1mm}
    \caption{Two-bus network representation.}
    \label{fig:twobus}
    \vspace{-4mm}
\end{figure}
Consider a two-bus network as shown in Fig.~\ref{fig:twobus}, where node A (slack/source) is connected to node B (load/generation point) via an impedance $Z_{AB}=R_{AB}+jX_{AB}$. Let $V_B$ be the complex voltage at node B, with net complex power injection $S_B=P_B+jQ_B$. Assuming the sending-end voltage $V_A$  is fixed at node A, the current flowing through the line is: 
\begin{equation}
    I_{AB}=(\frac{P_B+jQ_B}{V_B})^*
    \label{eq:cur}
\end{equation}

Applying Ohm’s law,  the complex voltage drop across the line is: 
\begin{equation}
    \Delta V_{AB}=I_{AB}.(R_{AB}+jX_{AB})
    \label{eq:drop}
\end{equation}
\noindent Substituting (\ref{eq:cur}) into (\ref{eq:drop}) and separating real and imaginary components, (\ref{eq:subcur}) is obtained. The first term in (\ref{eq:subcur}) is a real term influencing voltage magnitude and the second term is an imaginary term affecting the phase angle. 
\begin{equation}
    \Delta V_{AB}=\frac{P_BR_{AB}+Q_BX_{AB}}{|V_B|}+j\frac{P_BX_{AB}-Q_BR_{AB}}{|V_B|}
    \label{eq:subcur}
\end{equation}

In LV networks where $R_{AB} \gg X_{AB}$, the real part governs the magnitude of voltage deviation. Ignoring the phase angle, the magnitude approximation is:
\begin{equation}
    |\Delta V_{AB}|\approx\frac{P_BR_{AB}+Q_BX_{AB}}{|V_B|}
    \label{eq:vd}
\end{equation}

Taking the Jacobian of (\ref{eq:vd}), the sensitivity of voltage to active and reactive power is obtained by taking the partial derivatives as in \eqref{eq:jac}:
\begin{equation}
    |\frac{\delta|\Delta V|}{\delta P_B}|=\frac{R_{AB}}{{|V_B}|}, |\frac{\delta|\Delta V|}{\delta Q_B}|=\frac{X_{AB}}{{|V_B}|} 
    \label{eq:jac}
\end{equation}

In resistive grids (LV networks) $\frac{R_{AB}}{X_{AB}} \gg 1$, the voltage magnitude is more sensitive to active power injections as expressed in (\ref{eq:f}):
\begin{equation}
    |\frac{\delta|\Delta V|}{\delta P_B}|>> |\frac{\delta|\Delta V|}{\delta Q_B}|
    \label{eq:f}
\end{equation}

Conversely, in inductive HV systems:
\begin{equation}
    |\frac{\delta|\Delta V|}{\delta Q_B}|>> 
    |\frac{\delta|\Delta V|}{\delta P_B}|
    \label{eq:ff}
\end{equation}
\noindent Eqs. \eqref{eq:f}-\eqref{eq:ff} highlight that voltage regulation is governed predominantly by active power in resistive grids and reactive power in inductive networks..


\subsection{Functions of SIs}
SIs provide both steady-state and dynamic grid-support function. Their flexibility in changing control modes allows dynamic regulation of active and reactive power, enabling real-time grid support without additional devices (e.g., centralized communication infrastructure). In addition to steady-state control, SIs are also equipped with dynamic functionalities such as fault ride-through, allowing them to remain connected and support the grid during voltage sags or frequency excursions. This section discusses Volt-VAR and Volt-Watt control modes of SI \cite{sigsf}. 

\subsubsection{Volt-VAR Control}
Volt-VAR control modulates the reactive power output $Q$ of the inverter as a function of local voltage $V$ at the point of common coupling (PCC). This is typically implemented via a piecewise linear droop characteristic as expressed as (\ref{vv}): 
\begin{align}\label{vv}
    Q(V) 
    =\begin{cases}
       Q_{max}, \quad if \quad V \leq V_1 \\
       d_{vv}(V-V_{ref}), \quad if \quad V_1 <V<V_2 \\
       -Q_{max}, \quad if \quad V \geq V_2
       \end{cases}
\end{align}
where $V_{ref}$ is the nominal voltage, $V_1$ and $V_2$ are the minimum and maximum voltage thresholds at the PCC, $Q_{max}$ is the maximum reactive power, and $d_{vv}$ is the volt-var droop gain. 

\subsubsection{Volt-Watt Control}
Volt-Watt control adjusts active power output $P$ based on the local voltage magnitude. It  curtails active power to reduce overvoltage issues, which are common under high PV generation and light load conditions.
The control curve is defined by (\ref{vw}):  
\begin{align}\label{vw}
    P(V) 
    =\begin{cases}
       P_{rated}, \quad if \quad V \leq V_{ref} \\
       P_{rated}-d_{vw}(V-V_{ref}), \quad if \quad V_{ref} <V<V_2 \\
       0, \quad if \quad V \geq V_2
       \end{cases}
\end{align}
where $P_{rated}$ is the rated power of the inverter and $d_{vw}$ is the volt-watt droop gain. 





\section{Testbench and SI Control Modes}\label{sec:system_modeling}
\begin{figure}[t]
    \centering
    \includegraphics[scale=0.13]{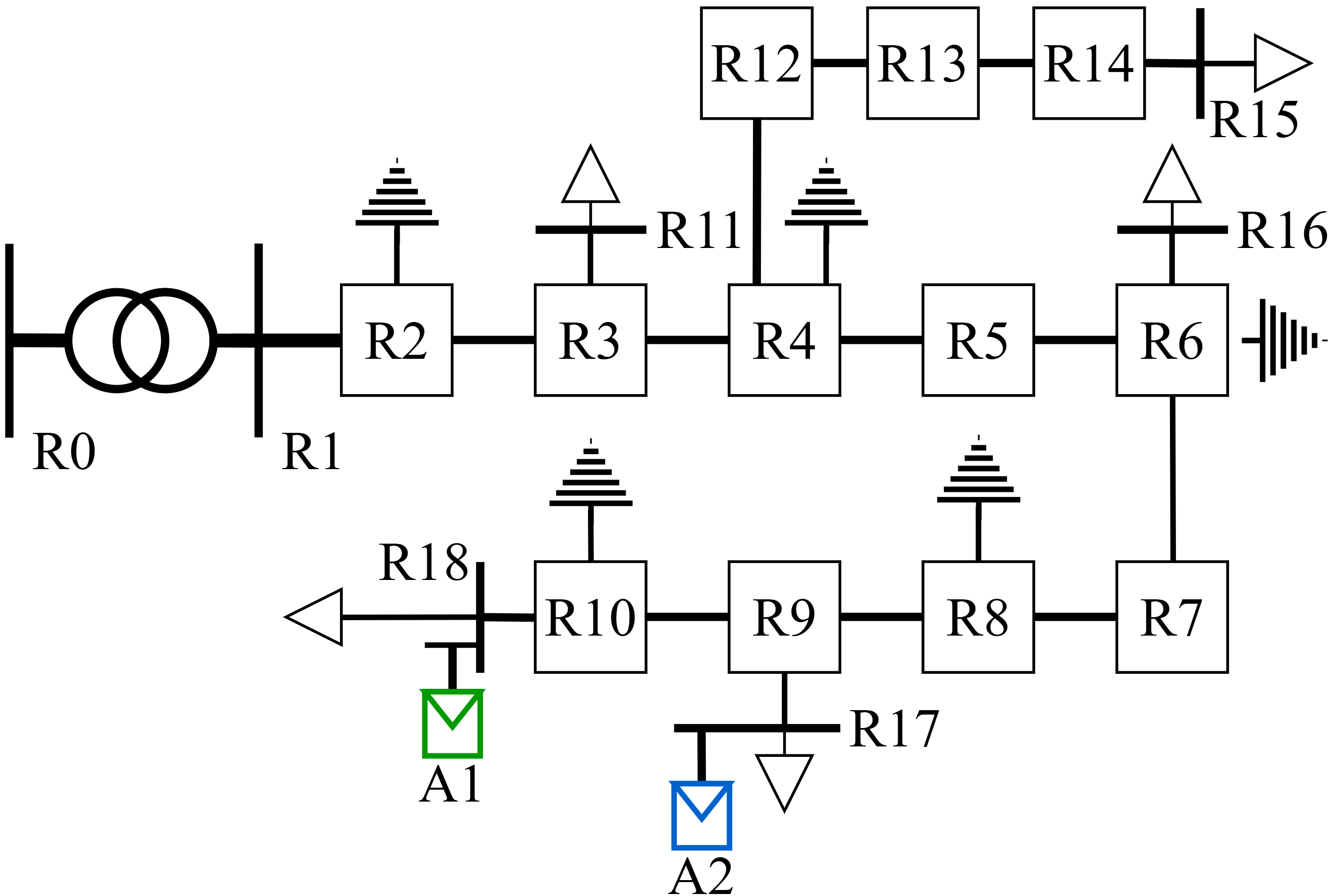}
    \caption{CIGRE LV residential subnetwork  with Smart Inverter Aggregators.}
    \label{fig:feeder}
    \vspace{-1mm}
\end{figure}

To evaluate the system-level impact of heterogeneous SI control strategies, a real-time simulation model of the CIGRE LV residential distribution network, as shown in Fig.~\ref{fig:feeder}. SI toolbox \cite{siopal}, which is developed using Opal-RT, is utilized to implement control dynamics of SI groups across various scenarios. 

The network model, based on the standardized CIGRE LV residential distribution network \cite{ctf} is modeled in MATLAB/Simulink and deployed to Opal-RT for real-time execution.
This CIGRE LV network represents a suburban radial feeder topology operating at 0.4 kV, 50 Hz, in a three-phase, four-wire configuration with multiple residential branches. This network  includes residential, industrial and commercial subnetworks. In this study, only the residential section of the network is considered, with loads modeled using aggregated active and reactive power values. The network and load specifications are provided in Table~\ref{tab:feeder_load}. 

\begin{table}[t]
    \centering
    \caption{CIGRE LV distribution network and load specifications.}
    \vspace{-1mm}
    \begin{tabular}{||c|c|c||}
        \hline \hline
        \multicolumn{3}{||c||}{\textbf{Network Specifications}} \\ \hline
        \textbf{Cable Type} & \textbf{Resistance ($\Omega$/km)} & \textbf{Inductance (mH/km)} \\ \hline
        UG1 & 0.287 & 0.5316 \\ \hline
        UG3 & 1.152 & 1.4579 \\ \hline
        \multicolumn{3}{||c||}{\textbf{Load Specifications}} \\ \hline
        \textbf{Node} & \textbf{Active Power (kW)} & \textbf{Reactive Power (kVAR)} \\ \hline
        R1 & 190 & 62.45 \\ \hline
        R11 & 14.25 & 4.68 \\ \hline
        R15 & 49.4 & 16.24 \\ \hline
        R16 & 52.25 & 17.17 \\ \hline
        R17 & 33.25 & 10.93 \\ \hline
        R18 & 44.65 & 14.68 \\ \hline \hline
    \end{tabular}
    \label{tab:feeder_load}
\end{table}

\begin{figure}[!t]
    \centering
    \subfloat[\centering ]{{\includegraphics[scale=0.99]{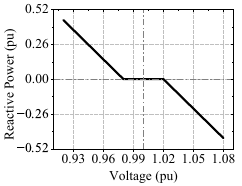}}
    \label{sub:vv}}
    \quad
    \subfloat[\centering ]{{\includegraphics[scale=0.99]{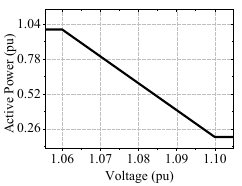}}
    \label{sub:vw}}
    \vspace{-1mm}
    \caption{Control curves of SIs: (a) Volt-VAR, (b) Volt-Watt.}%
    \label{fig:curves}%
    \vspace{-4mm}
\end{figure}

Two SI groups, each consisting of five 10 kVA inverters, are deployed at nodes R17 and R18 of the CIGRE LV residential network. These groups are hereafter referred to as Aggregator 1 (A1) and Aggregator 2 (A2), respectively. Fig.~\ref{fig:feeder} shows the single-line diagram of the network with SIs and the placements. Within each aggregator, all SIs operate under the same control mode to facilitate a systematic evaluation of how one mode influences the performance of another at the aggregator level. This deployment facilitates the analysis of both localized and system-wide voltage regulation effects of different control strategies of SIs. The implemented Volt-VAR and Volt-Watt control characteristics are shown in Fig.~\ref{sub:vv} and Fig.~\ref{sub:vw}, respectively. 



\section{Simulation Results}\label{sec:results}
To assess the impact of heterogeneous control modes of SIs in A2 on the operation of SIs in A1, real-time simulations are conducted on the CIGRE LV distribution network. SIs in A1 operate in the same control mode (CPF in this case) throughout all simulations, and the control modes of SIs in A2 are varied across three modes:  CPF (M1), Volt-VAR (M2), and Volt-Watt (M3). In CPF mode, the SIs maintain a power factor of 0.95 (lagging).
Simulations are carried out under the following timeline of events:
\begin{enumerate}
    \item \(0<t<5~s\): Normal operation.
    \item \(5<t<10~s\): Irradiance drops to 50\% of its initial value.
    \item \(10<t<13~s\): Irradiance is restored.
    \item \(13<t<17~s\): Grid voltage is increased to 1.2~pu.
    \item \(17<t<20~s\): Grid voltage is returned to nominal 1~pu.
\end{enumerate}

The impact of of A2’s control modes  is evlauated  by measuring active, reactive power, power factor and voltage at the PCC of A1 and A2 under both resistive ($\uparrow \frac{R}{X}$) and inductive ($\uparrow \frac{X}{R}$) grids.  


\subsection{Impact Assessment of Control Modes in Resistive Grid}

\begin{figure}[!t]
    \centering
    \subfloat[\centering ]{{\includegraphics[scale=0.90]{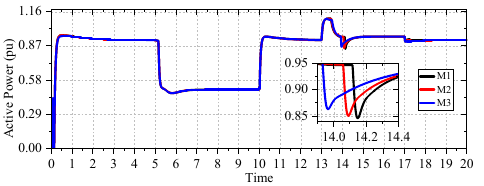}}
    \label{sub:rpa1}}
    \quad
    \subfloat[\centering ]{{\includegraphics[scale=0.90]{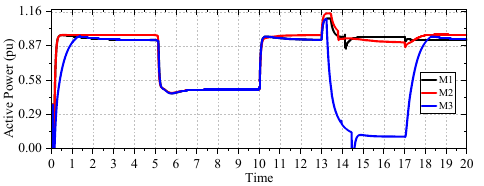}}
    \label{sub:rpa2}}
    \vspace{-1mm}
    \caption{Active power at PCC in resistive grid of aggregators: (a) A1, (b) A2.}%
    \label{fig:rpa}%
    \vspace{-4mm}
\end{figure}

Figs.~\ref{sub:rpa1} and \ref{sub:rpa2} show that changing the control mode of the SIs in A2 from CPF to Volt-VAR or Volt-Watt significantly affects the active power response of SIs in A1, particularly during event 4 (voltage rise at 13-17~s). Compared to the CPF base case (black line), operating A2 in Volt-VAR mode (red line) results in a slightly improved transient response at A1, with reduced active power dip and delay. The improvement is more pronounced when A2 operates in Volt-Watt mode (blue line), where the active power transients at A1 are significantly mitigated, indicating improved voltage regulation and reduced upstream disturbance.


This behavior is attributed to the influence of A2’s control actions on A1’s voltage profile via shared network impedance. In resistive grids, when A2 operates in Volt-Watt mode, it rapidly curtails active power in response to overvoltage, reducing local voltage and consequently mitigating the upstream voltage seen at A1. This leads to a smaller and faster transient response in A1’s power output. In contrast,Volt-VAR mode, which is less effective in resistive networks, results in a slightly larger dip and delay compared to Volt-Watt.

\begin{figure}[!t]
    \centering
    \subfloat[\centering ]{{\includegraphics[scale=0.90]{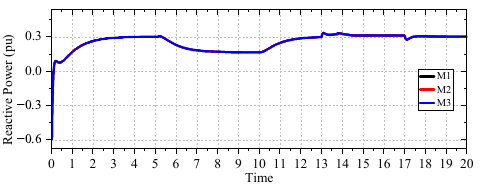}}
    \label{sub:rqa1}}
    \quad
    \subfloat[\centering ]{{\includegraphics[scale=0.90]{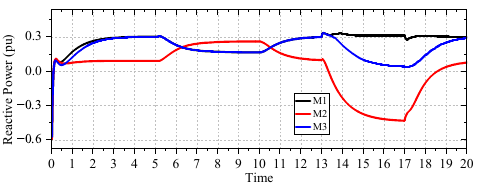}}
    \label{sub:rqa2}}
    \vspace{-1mm}
    \caption{Reactive power at PCC in resistive grid of aggregators: (a) A1, (b) A2.}%
    \label{fig:rqa}%
    \vspace{-4.8mm}
\end{figure}

\begin{figure}[!t]
    \centering
    \subfloat[\centering ]{{\includegraphics[scale=0.90]{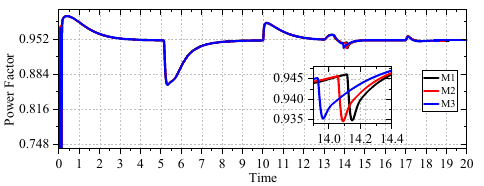}}
    \label{sub:rpfa1}}
    \quad
    \subfloat[\centering ]{{\includegraphics[scale=0.90]{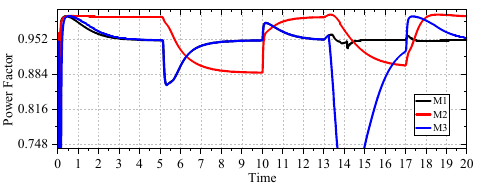}}
    \label{sub:rpfa2}}
    \vspace{-1mm}
    \caption{Power factor at PCC in resistive grid of aggregators: (a) A1, (b) A2.}%
    \label{fig:rpfa}%
    \vspace{-4mm}
\end{figure}

Fig.~\ref{sub:rqa1} and \ref{sub:rqa2} show that the reactive power output of A1 remains largely unaffected by A2's control mode. Correspondingly, the power factor behavior of A1 aligns with active power trends during transients as seen in Fig.~\ref{sub:rpfa1} and \ref{sub:rpfa2}.

\begin{figure}[!t]
    \centering
    \vspace{-1.5mm}
    \subfloat[\centering ]{{\includegraphics[scale=0.90]{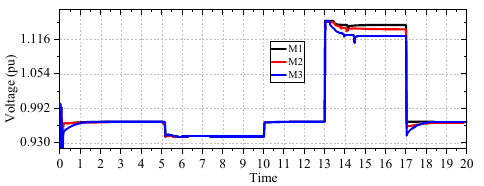}}
    \label{sub:rva1}}
    \quad
    \subfloat[\centering ]{{\includegraphics[scale=0.90]{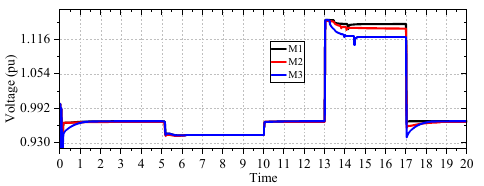}}
    \label{sub:rva2}}
    \vspace{-1mm}
    \caption{Voltage at PCC in resistive grid of aggregators: (a) A1, (b) A2.}%
    \label{fig:rva}%
    \vspace{-4.8mm}
\end{figure}

Fig.~\ref{sub:rva1} and \ref{sub:rva2} show that the voltage rise during event 4 is high when A2 operates in CPF mode, as no regulation occurs. Volt-VAR mode reduces the voltage via reactive power absorption but is limited by low Q-V sensitivity in this resistive grid. Volt-Watt control mode achieves the most effective voltage suppression through active power curtailment, both primarily at A2 and to a lesser extent upstream at A1.

\subsection{Impact Assessment of Control Modes in Inductive Grid}
To analyze behavior in inductive grids, the line impedance of the CIGRE LV distribution network (which is typically resistive in nature)  is modified by setting the inductance to five times the resistance. This  configuration emphasizes the role of active/reactive power in voltage regulation due to higher $X/R$ ratios.

\begin{figure}[!t]
    \centering
    \subfloat[\centering ]{{\includegraphics[scale=0.90]{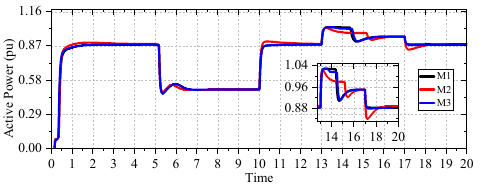}}
    \label{sub:ipa1}}
    \quad
    \subfloat[\centering ]{{\includegraphics[scale=0.90]{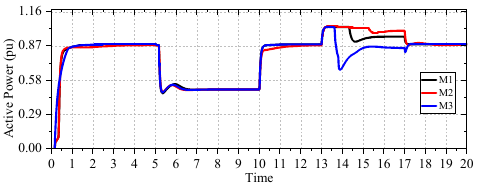}}
    \label{sub:ipa2}}
    \vspace{-1mm}
    \caption{Active power at PCC in inductive grid of aggregators: (a) A1, (b) A2.}%
    \label{fig:ipa}%
    \vspace{-4mm}
\end{figure}

Fig.~\ref{sub:ipa1} and \ref{sub:ipa2} show that the active power response of SIs in A1 is nearly identical when SIs in A2 operate in CPF and Volt-Watt modes. However, Volt-VAR operation at A2 induces a noticeable delayed response in A1’s active power. 
This is because the CPF and Volt-Watt control modes at A2 do not significantly alter reactive power, leading to minimal influence on upstream voltage and, consequently, negligible impact on A1’s active power.

\begin{figure}[!t]
    \centering
    \subfloat[\centering ]{{\includegraphics[scale=0.90]{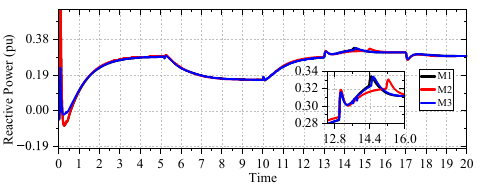}}
    \label{sub:iqa1}}
    \quad
    \subfloat[\centering ]{{\includegraphics[scale=0.90]{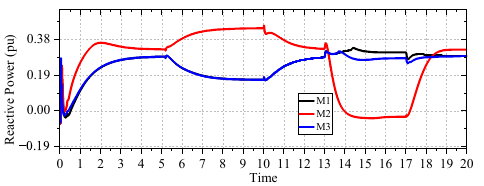}}
    \label{sub:iqa2}}
    \vspace{-1mm}
    \caption{Reactive power at PCC in inductive grid of aggregators:(a)A1 and (b) A2.}%
    \label{fig:iqa}%
    \vspace{-4mm}
\end{figure}

\begin{figure}[!t]
    \centering
    \subfloat[\centering ]{{\includegraphics[scale=0.90]{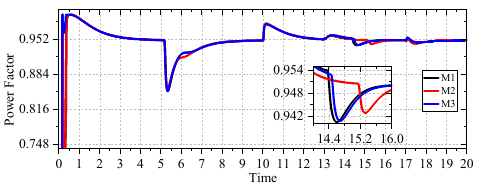}}
    \label{sub:ipfa1}}
    \quad
    \subfloat[\centering ]{{\includegraphics[scale=0.90]{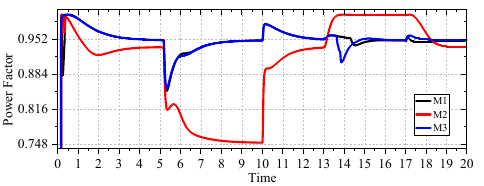}}
    \label{sub:ipfa2}}
    \vspace{-1mm}
    \caption{Power factor at PCC in inductive grid of aggregators: (a) A1, (b) A2.}%
    \label{fig:ipfa}%
    \vspace{-5mm}
\end{figure}

In contrast, the Volt-VAR mode of A2 adjusts reactive power dynamically in response to voltage changes. However, due to controller ramp rate limits, Volt-VAR mode exhibits a delayed impact on voltage, which indirectly causes a delay in A1’s active power response. The same scenario is observed in the reactive power graphs for both A1 and A2; see  Fig.~\ref{sub:iqa1} and \ref{sub:iqa2}. The power factor profiles of A1 are affected in the same way as the active power, as observed in Fig.~\ref{sub:ipfa1} and \ref{sub:ipfa2}. 

\begin{figure}[!t]
    \centering
    \subfloat[\centering ]{{\includegraphics[scale=0.90]{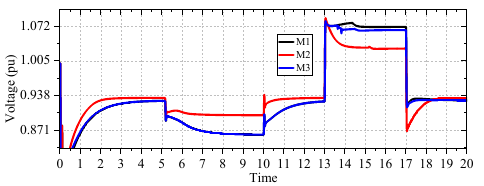}}
    \label{sub:iva1}}
    \quad
    \subfloat[\centering ]{{\includegraphics[scale=0.90]{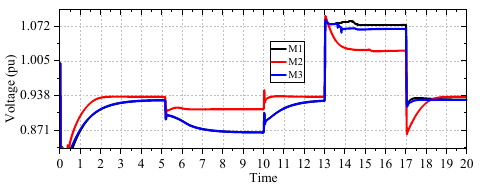}}
    \label{sub:iva2}}
    \vspace{-1mm}
    \caption{Voltage at PCC in inductive grid of aggregators: (a) A1, (b) A2.}%
    \label{fig:iva}%
    \vspace{-5mm}
\end{figure}

Fig.~\ref{sub:iva1} and \ref{sub:iva2} reveal that the voltage rise during event 4 is again highest under CPF mode due to the absence of regulation. In this case, Volt-VAR mode is highly effective, absorbing reactive power to significantly reduce voltage at A2, leveraging strong Q-V sensitivity in inductive networks. Volt-Watt mode also mitigates voltage rise, but its effect is less compared to Volt-VAR.

\section{Conclusion}\label{sec:conclusion}
This study analyzes the influence of control modes of SIs on other SI groups within a LV distribution network using a real-time SI testbench. The influence of Volt-VAR and Volt-Watt modes is evaluated under both resistive and inductive grid conditions. In both scenarios, the control mode deployed at one aggregator influenced the local voltage and, consequently, the participation of neighboring SIs in voltage regulation. 
The results, obtained though real-time simulations using Opal-RT, demonstrate that in resistive grids Volt-Watt  at one aggregator significantly influenced upstream voltage and active power dynamics of nearby SIs. In contrast, in inductive grids, Volt-VAR is more effective in regulating local voltage without substantial system-wide impact. 
These findings highlight the importance of deployment of SI control modes. The grid's impedance characteristics should guide control mode selection to improve system-wide voltage stability and operational resilience,  especially in grids with high renewable penetration.

\section*{Acknowledgments}
This publication is based upon work supported by KAUST under Award No. ORFS-CRG11-2022-5021 as well as KAUST - Center of Excellence for Renewable Energy and Storage Technologies (CREST), under award number 5937.

\bibliographystyle{IEEEtran}
\bibliography{References}

\end{document}